\documentclass[12pt]{article}

\usepackage[margin=1in]{geometry}
\usepackage{setspace}
\usepackage{hyperref}
\usepackage{siunitx}
\usepackage{amsmath}
\usepackage{natbib}
\usepackage{titlesec}
\usepackage{tikz}
\usepackage{graphicx}
\usepackage{subcaption}
\usetikzlibrary{arrows.meta, positioning}

\titleformat{\section}{\normalfont\Large\bfseries}{\thesection}{1em}{}
\titleformat{\subsection}{\normalfont\large\bfseries}{\thesubsection}{1em}{}

\title{The Quiet Skies Report\\
\textit{\large A Primer on Protecting Radio Astronomy in the Age of Satellite Mega-Constellations}}

\author{Gregory Hellbourg\\California Institute of Technology\\ghellbourg@astro.caltech.edu}

\date{}

\begin{document}

\maketitle

\begin{abstract}
The rapid expansion of satellite constellations is transforming the radio-frequency environment around the Earth. At the same time, radio astronomy is entering a new era of sensitivity and survey capability, requiring unprecedented control of interference. This primer introduces satellite operators, engineers, spectrum managers and policy makers to the basic concepts of radio astronomy, explains why the discipline is uniquely vulnerable to interference, and outlines the regulatory and practical tools available to manage coexistence.
\end{abstract}

\tableofcontents

\clearpage

\section*{Executive Summary}
\addcontentsline{toc}{section}{Executive Summary}

Satellite operators increasingly share the sky with some of the most sensitive scientific instruments ever built. Radio telescopes detect natural radio waves from astronomical sources at power levels many orders of magnitude weaker than those involved in communication links. In some bands, the harmful interference threshold for a radio telescope corresponds to a received power of order \SI{-250}{dBW} or lower at its input. For comparison, this is roughly equivalent to detecting the signal from a mobile phone tens of millions of kilometers away. Despite the technical sophistication of modern payloads, small amounts of unwanted radiation, whether intentional out-of-band emissions or unintended electromagnetic leakage from spacecraft subsystems, can accumulate in the data produced by these instruments and compromise their scientific value.

Satellite operators should care about radio astronomy for several reasons. First, there are clear global regulatory obligations. The Radio Regulations of the International Telecommunication Union (ITU) recognize the Radio Astronomy Service (RAS) as a legitimate and particularly vulnerable user of the spectrum. They define protected bands, footnotes and protection criteria that administrations are expected to respect. Second, there are economic and reputational risks. Harmful interference to major observatories can result in regulatory scrutiny, delays to licensing, and negative public attention, particularly where the affected science relates to widely appreciated topics such as black holes, planetary discovery or cosmology. Third, there are positive incentives. Demonstrating serious attention to coexistence issues strengthens an operator's position as a responsible actor in the broader space sustainability agenda, a topic increasingly relevant to United Nations bodies and national regulators.

Collaboration brings tangible benefits. When operators and observatories communicate early, compatibility issues can often be identified while there is still flexibility in system design. Shared analysis of aggregate interference, joint test campaigns, and agreed procedures for reporting and investigating unexpected interference episodes can prevent small problems from becoming public disputes. Good cooperation also improves the quality of scientific and engineering data available to all parties, leading to more robust designs.

From a practical perspective, satellite operators can think in terms of a checklist of actions. They can ensure that their systems respect conservative limits on power spectral density, not only in the allocated bands, but also out-of-band, taking into account filter roll-off and intermodulation products. They can pay attention to antenna patterns and sidelobe levels over known radio observatories, including the possibility of reducing power or avoiding certain beam configurations when spacecraft pass over sensitive sites. They can control duty cycles and burst scheduling so that particularly intense transmissions do not coincide with time-critical astronomical observations. They can characterize and minimize unintended electromagnetic radiation from power electronics and digital subsystems. Finally, they can maintain clear lines of communication with observatories and regulators for the reporting and joint investigation of interference events, including the sharing of accurate and timely satellite ephemeris so that observatories can anticipate when satellites will cross their fields of view and adjust sensitive observations accordingly.

The rest of this primer provides the background needed to understand why these steps matter, how radio astronomy works, what kinds of interference are most problematic, and how international and national regulations frame the protection of radio astronomy. It concludes with a discussion of future trends, including the next generation of instruments such as the Deep Synoptic Array (DSA-2000) and the Square Kilometre Array (SKA), and new policy concepts such as ``dark and quiet skies'' and electromagnetic environmental protection.

\clearpage

\section{What is Radio Astronomy?}

\subsection{Astronomical signals and wavelengths}

Radio astronomy is the study of the Universe using radio waves, typically at frequencies from a few megahertz to hundreds of gigahertz. These waves occupy the same electromagnetic spectrum as visible light, infrared radiation, and X-rays, but they differ only in frequency and wavelength. At radio wavelengths, different physical processes dominate. Cold hydrogen gas in galaxies emits a characteristic spectral line at \SI{1420}{\mega\hertz}, a benchmark transition used to map the structure and motion of galaxies. Molecules such as hydroxyl, water and ammonia emit at other specific frequencies, revealing the chemistry and conditions in star-forming regions. Synchrotron radiation from relativistic electrons gives insight into shocks, jets and magnetic fields. Pulsars, which are rotating neutron stars, emit beams of coherent radio emission that sweep across the Earth, producing precise pulses that can be used as cosmic clocks.

Compared to optical astronomy, radio astronomy often observes regions that are invisible at shorter wavelengths. Dust that blocks starlight is largely transparent at centimeter and meter wavelengths. Radio telescopes can see into the center of our Galaxy and into dense molecular clouds. Radio observations can also be carried out by day and by night, in a range of weather conditions. These features make radio astronomy complementary to optical, infrared, and high-energy astronomy.

Radio astronomy is uniquely fragile because of the faintness and spectral purity of the signals it seeks. A useful unit for measuring source strength is the Jansky (Jy), defined as
\[
  1~\mathrm{Jy} = 10^{-26}~\mathrm{W\,m^{-2}\,Hz^{-1}}.
\]
To put this into context, a typical satellite downlink or cellular transmission corresponds to flux densities millions of billions of times larger—often around $10^{-13}$ to $10^{-16}~\mathrm{W\,m^{-2}\,Hz^{-1}}$ at the surface of the Earth. Many sources of scientific interest have flux densities of millijanskys or less, corresponding to \(10^{-29}\,\mathrm{W\,m^{-2}\,Hz^{-1}}\) or below, meaning that radio telescopes routinely work with power levels more than ten trillion times weaker than those used in everyday communication links. When such a weak signal is collected by a telescope of finite area and integrated over a limited bandwidth, the resulting power at the receiver input is often tens of decibels (10s to 10,000s times) below the thermal noise generated within the receiver itself. Radio astronomers use long integrations, correlation across many antennas, and sophisticated calibration to recover these signals statistically. Any additional interference, even if weak and sporadic, can bias this process.

\subsection{History and societal value}

The origins of radio astronomy lie in engineering. In the early 1930s, Karl Jansky, working for Bell Telephone Laboratories, was tasked with identifying sources of static that interfered with shortwave radio links. Using a rotating antenna, he detected a mysterious hiss whose position in the sky shifted with the sidereal day rather than the solar day, indicating a cosmic origin \citep{jansky1933}. Subsequent work by Grote Reber and others mapped the radio emission from the Milky Way and discovered discrete radio sources, including remnants of supernova explosions and active galaxies.

In the decades that followed, radio astronomers made a series of discoveries with profound scientific and cultural impact. The identification of radio galaxies and quasars revealed the presence of extremely energetic processes in distant galaxies. The detection of the \SI{1420}{\mega\hertz} hydrogen line allowed detailed mapping of the spiral structure and dynamics of the Milky Way. In 1965, Arno Penzias and Robert Wilson discovered the cosmic microwave background radiation while investigating excess noise in a horn antenna built for satellite communication experiments, providing strong support for the Big Bang model. The discovery of pulsars in 1967 opened a new window on compact objects and dense matter. More recently, radio observations have contributed to the first image of a black hole's shadow by the Event Horizon Telescope, a global very long baseline interferometry (VLBI) array observing at millimeter wavelengths.

Radio astronomy also plays a central role in planetary science, cosmology and astrometry. Planetary radar and radio observations probe the surfaces and atmospheres of Solar System bodies. Measurements of distant galaxies and the intergalactic medium inform models of cosmic structure formation and dark energy. VLBI networks provide highly accurate reference frames and monitor the rotation and deformation of the Earth, underpinning navigation and geodetic applications. Many of these contributions have practical as well as scientific value.

The societal value of radio astronomy extends far beyond pure knowledge. The field has driven major technological developments in low-noise receivers, cryogenics, digital signal processing, spectrum analysis and interferometric methods. Several technologies now taken for granted have roots in radio astronomical research: early work on low-noise microwave electronics and spread-spectrum techniques contributed to the foundation of Wi-Fi, and advances in image reconstruction algorithms influenced the development of magnetic resonance imaging (MRI). Techniques pioneered in radio astronomy have shaped wireless communications, medical imaging and remote sensing, and continue to inform modern sensor and signal-processing architectures.

Radio astronomy has also become a pillar of global geodesy and navigation. Very Long Baseline Interferometry (VLBI), initially developed to image radio quasars, now forms the backbone of the International VLBI Service for Geodesy and Astrometry (IVS). By observing distant quasars that define the International Celestial Reference Frame (ICRF), geodetic VLBI arrays measure Earth rotation, nutation, polar motion and tectonic plate drift with milliarcsecond precision. These measurements tie together the terrestrial and celestial reference frames, and are essential inputs to maintaining the long-term accuracy of Global Navigation Satellite Systems (GNSS) such as the Global Positioning System (GPS), and its international counterparts Galileo, GLONASS, and BeiDou. Without VLBI, GNSS timing stability, orbit determination, and global position consistency would degrade over time. In this sense, radio astronomy is not only a scientific endeavor but also a critical component of the world’s positioning, navigation and timing infrastructure.

Training in radio astronomy produces highly skilled engineers and scientists who frequently transition into the broader space, telecommunications, computing and instrumentation sectors. The field’s contributions, ranging from deep cosmological insights to enabling modern navigation systems, demonstrate the wide societal and technological benefits that arise from protecting the radio spectrum for passive scientific use.

\subsection{Hot topics today}

Contemporary radio astronomy is characterized by a set of ``hot topics'' that illustrate the breadth and ambition of the field. Fast radio bursts (FRBs) are millisecond-duration flashes of radio waves with high dispersion, indicating that they have traveled through large columns of ionized gas, most likely from cosmological distances far beyond our own Milky Way galaxy. They are a focus of intensive observational and theoretical work and may serve as probes of the baryon content of the Universe, provided that interference can be controlled at the required time and frequency resolution.

Neutral hydrogen mapping remains a central goal. Large surveys of the \SI{1420}{\mega\hertz} line in emission and absorption trace the distribution and kinematics of gas in galaxies and can be used to measure large-scale structure and baryon acoustic oscillations. Low-frequency experiments seek to detect the faint signatures of hydrogen in the early Universe, during the epochs of recombination and reionization. These observations place extreme demands on spectral purity and interference control.

Space weather and solar radio physics are another active area. Radio telescopes monitor solar flares, coronal mass ejections, and radio bursts that can affect satellite operations, GNSS signals, and power grids. In this sense, radio astronomy provides services directly relevant to the space industry. Searches for technosignatures, often associated with the broader term Search for Extra-Terrestrial Intelligence (SETI), use radio telescopes to look for narrowband or structured signals that might indicate artificial transmitters around other stars. Finally, deep surveys with instruments such as the Deep Synoptic Array (DSA-2000) or the Square Kilometre Array (SKA) are designed to map large portions of the sky with unprecedented sensitivity and time resolution, generating data that will be mined for many types of transient and persistent phenomena.

All of these topics rely on stable, low-interference observing conditions. Some target narrow spectral lines, and others use very wide bands but demand particularly clean behavior at certain frequencies. In each case, uncontrolled emissions from satellites can have a disproportionate impact.

\section{How Radio Astronomy Works}

\subsection{Receiving extremely weak signals}

The defining feature of radio astronomy is its focus on extremely weak signals. The power \(P\) received from a source with flux density \(S\) (in \(\mathrm{W\,m^{-2}\,Hz^{-1}}\)) by a telescope with effective collecting area \(A_{\mathrm{eff}}\) over a bandwidth \(\Delta \nu\) is approximately
\[
  P \approx S \, A_{\mathrm{eff}} \, \Delta \nu.
\]
For a source of one Jansky observed with a telescope of effective area, say, \(10^{3}\,\mathrm{m^{2}}\) over a \SI{1}{\mega\hertz} bandwidth, this corresponds to
\[
P \sim 10^{-26} \times 10^{3} \times 10^{6}~\mathrm{W} = 10^{-17}~\mathrm{W},
\]
or \SI{-170}{dBW}. To give a more intuitive sense of scale, a single mobile phone transmitting at typical power from an aircraft \(\sim\)10~km above the telescope would deliver a received power of order \(10^{-12}\,\mathrm{W}\) at the antenna, about one hundred thousand times stronger than the signal from this one Jansky astronomical source. Radio telescopes are therefore routinely working with signals many orders of magnitude weaker than those generated by everyday communication devices, even when those devices are very far away. Many astronomical sources are far fainter than one Jansky (often at the millijansky or microjansky level), so the received power is correspondingly lower.

The noise in a radio receiver is often described in terms of a system temperature \(T_{\mathrm{sys}}\), which represents the combined effect of receiver electronics, spillover and sky background. In radio engineering, temperature is related to noise power through the relation \(P = k_{\mathrm{B}} T_{\mathrm{sys}} \Delta\nu\), where \(k_{\mathrm{B}}\) is Boltzmann’s constant. A higher effective temperature therefore corresponds directly to a higher noise power in watts.
A typical modern receiver may have \(T_{\mathrm{sys}}\) of 20–50~K at centimeter wavelengths (roughly 1--30~GHz), comparable to the temperature of liquid nitrogen. Against such a quiet background, the astronomical signal of interest is usually tens of decibels below the thermal noise already present at the receiver input.

The random fluctuations in measured power decrease with increasing bandwidth and integration time. For a simple total-power measurement with one polarization (most radio telescopes sample two polarizations at once), the radiometer equation gives the root-mean-square uncertainty in the measured antenna temperature:
\[
  \sigma_T \approx \frac{T_{\mathrm{sys}}}{\sqrt{\Delta \nu \, \tau}},
\]
where \(\tau\) the integration time. Expressed in terms of flux density, a similar relation shows that the minimum detectable flux density decreases as \(1/\sqrt{\Delta \nu \tau}\). In an interferometric array with many antennas and baselines, the effective sensitivity can improve further, roughly as the square root of the number of independent baselines. For example, an array with 2000 antennas produces nearly two million independent baselines, improving sensitivity by more than three orders of magnitude over a single dish.

These relationships highlight why radio astronomy uses wide bandwidths and long integrations. They also explain why even very small additional contributions to the noise budget, including low-level interference that is not obviously visible in a single short snapshot, can accumulate to become scientifically significant. A signal that is completely invisible in one second of data can become a clear contaminant after hours of averaging, much as a faint background hum in a quiet room becomes noticeable when listened to carefully over a long period.

\subsection{Instrumentation}

\begin{figure}[t]
    \centering
    \begin{subfigure}[b]{0.44\textwidth}
        \centering
        \includegraphics[width=\textwidth]{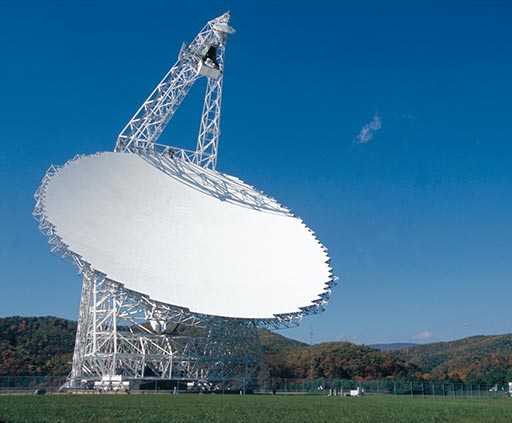}
        \caption{Green Bank Telescope, WV, USA.}
        \label{fig:gbt}
    \end{subfigure}
    \begin{subfigure}[b]{0.55\textwidth}
        \centering
        \includegraphics[width=\textwidth]{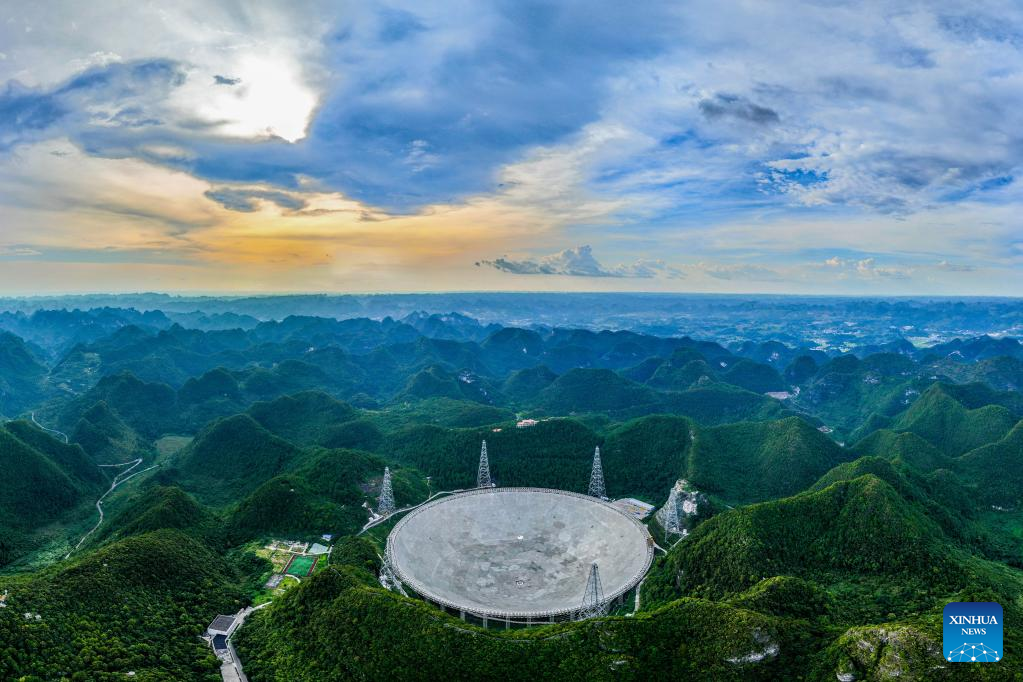}
        \caption{FAST Telescope, China.}
        \label{fig:fast}
    \end{subfigure}

    \caption{Examples of single-dish radio telescopes. Left: The Green Bank Telescope (GBT) in West Virginia, USA. With a fully steerable 100-meter diameter reflector, it is the world’s largest fully movable radio telescope (credit : NSF/NRAO). Right: The Five-hundred-meter Aperture Spherical Telescope (FAST) in Guizhou, China. FAST has an illuminated aperture of approximately 300 meters within a 500-meter fixed spherical dish. Although the structure itself is stationary, the telescope achieves sky coverage by moving its feed cabin and actively deforming sections of its reflector to form a steerable paraboloid (credit : Xinhua/Ou Dongqu).}
    \label{fig:single_dish}
\end{figure}

A radio telescope typically consists of three components: an antenna, a receiver chain, and a digital back end. In a single-dish system like the ones depicted in figure \ref{fig:single_dish}, a parabolic reflector focuses incoming radio waves onto a feed horn. The feed couples the signal into a low-noise amplifier (LNA), often cooled cryogenically so that the amplifier's own thermal noise does not overwhelm the astronomical signal. After amplification, the signal passes through filters that define the band of interest and suppress strong transmissions outside it. For transport over long distances (from the antenna to a central control building), the signal is frequently converted from radio frequency to light using RF-over-fiber links, which avoid the losses and electromagnetic contamination associated with long coaxial cables.

At the central facility, the signal is downconverted in frequency, and digitized by analogue-to-digital converters (ADCs) operating at high sampling rates and with sufficient dynamic range to retain the scientifically relevant information. Digital back ends then perform channelization (the decomposition of the signal into narrow frequency bins), correlation (multiplication of the signal with itself to recover the instantaneous power received at an antenna, or from different antennas to measure spatial coherence), beamforming (the electronic steering of beams without physically moving antennas), and integration over time, much like a long exposure in photography accumulates light to reveal a faint image.

\begin{figure}[t]
    \centering
    \begin{subfigure}[t]{0.48\textwidth}
        \centering
        \includegraphics[height=0.28\textheight,keepaspectratio]{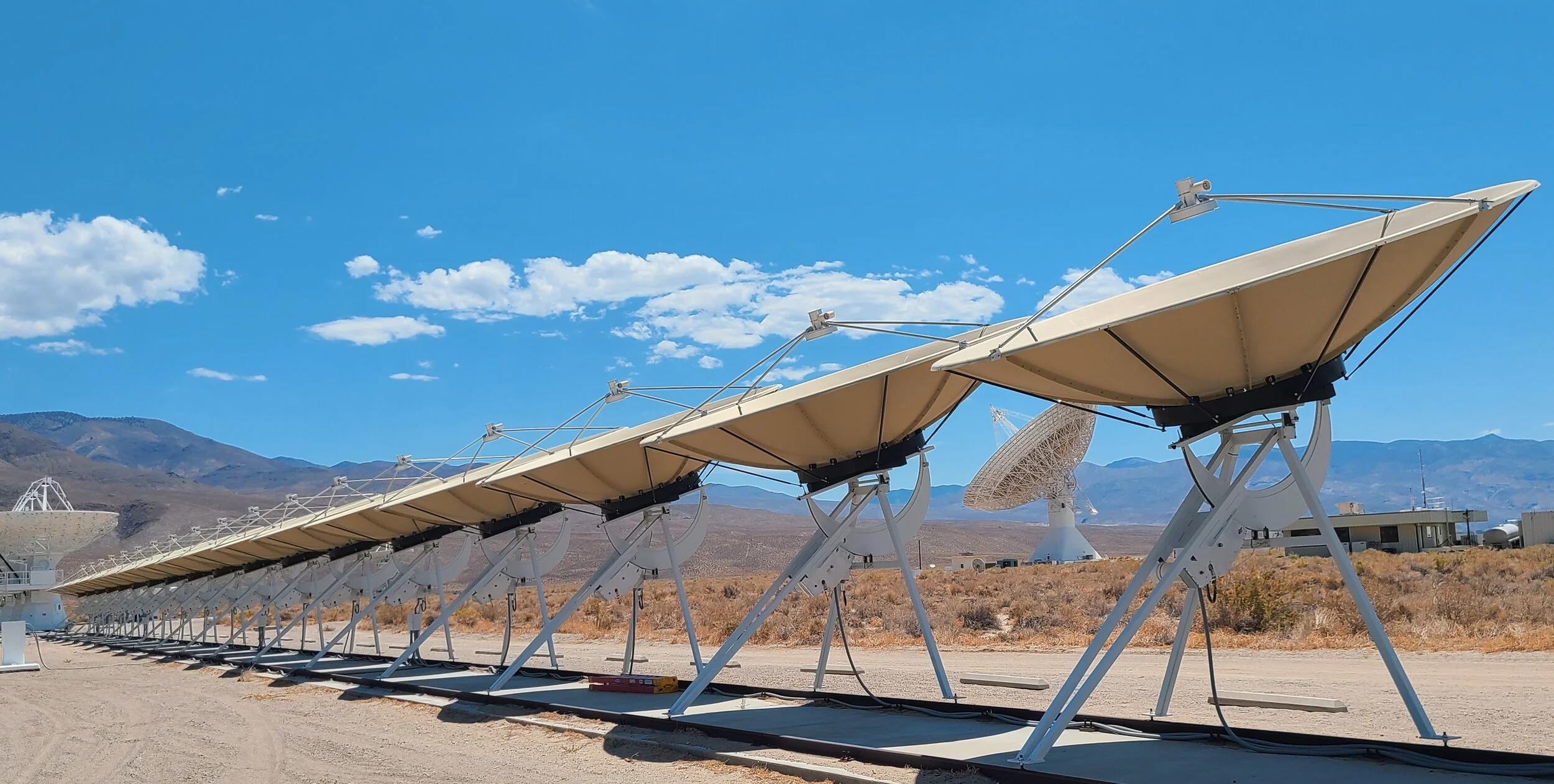}
        \caption{The Caltech Deep Synoptic Array (DSA-110), CA, USA.}
        \label{fig:dsa110}
    \end{subfigure}\hfill
    \begin{subfigure}[t]{0.48\textwidth}
        \centering
        \includegraphics[height=0.28\textheight,keepaspectratio]{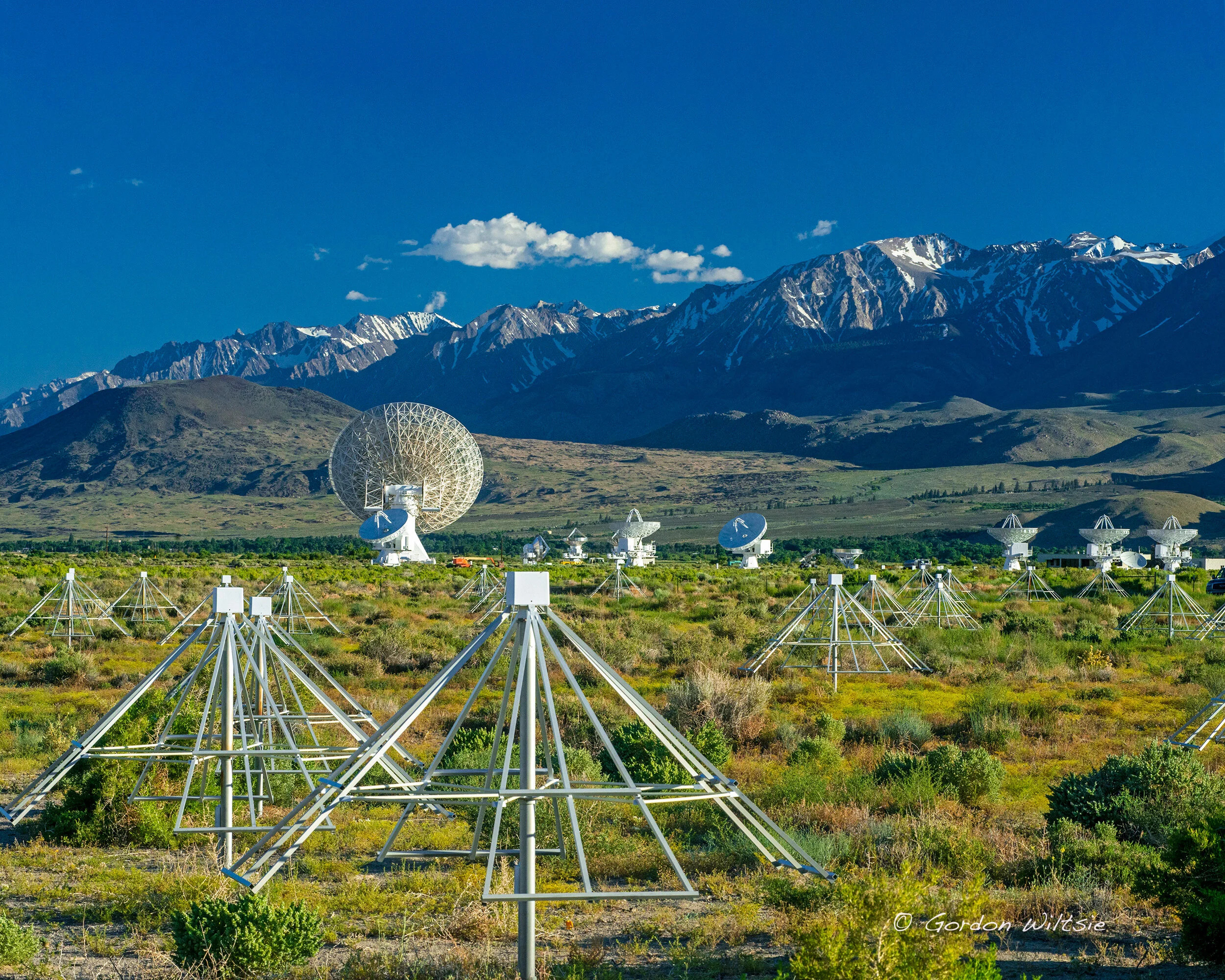}
        \caption{The Caltech Owens Valley Radio Observatory Long Wavelength Array (OVRO-LWA), CA, USA.}
        \label{fig:lwa}
    \end{subfigure}

    \caption{Examples of radio interferometers. Left: The Deep Synoptic Array 110 (DSA-110) at the Owens Valley Radio Observatory (OVRO) in California, USA. The instrument consists of 110 parabolic antennas, each 4.5 m in diameter, operating primarily in the 1.28–1.53 GHz band. Its primary mission is the rapid detection and precise localization of fast radio bursts (FRBs) (credit : G. Hallinan). Right: The Owens Valley Radio Observatory Long Wavelength Array (OVRO-LWA), composed of 352 dual-polarization dipole antennas with nearly omnidirectional response. The array operates over 20–80 MHz, enabling full-sky imaging at high cadence as well as beamformed observations for transient and space-weather science (credit : G. Wiltsie).}
    \label{fig:interferometers}
\end{figure}

Interferometers, like the ones depicted in figure \ref{fig:interferometers}, consist of many such signal chains, each attached to its own antenna. The digitized signals from all antennas are combined to form cross-correlations for every pair of antennas. These correlations measure different spatial Fourier components of the sky brightness distribution. By combining these measurements over time and frequency, astronomers reconstruct images using aperture synthesis techniques \citep{thompson2017interferometry}. Phased arrays, in which signals from many small elements are added with appropriate phase shifts, provide electronically steerable beams and are particularly powerful for rapid survey observations and for adaptive interference mitigation.

From the standpoint of coexistence with satellites and other active services, several aspects of this instrumentation are crucial. The front ends are optimized for low noise, not for high power handling. Strong interfering signals can push LNAs, mixers or ADCs into non-linear regimes, generating harmonics and intermodulation products that were not present in the original sky signal. These spurious products can contaminate a much wider frequency range than the interfering transmission itself. This behavior is analogous to an overexposed camera image, where a bright source does not merely saturate the pixels at its location, but can cause streaks and artifacts across the entire image.

Filters also provide only finite rejection. Even high-quality filters have limited roll-off, meaning that out-of-band emissions or spurious signals from satellites can leak into nominally protected bands. Similarly, ADCs have finite bit depth and a fixed full-scale range. If an interfering signal occupies too much of this range, the quantization noise affecting the remaining (weaker) astronomy signal increases. This is analogous to a digital camera with limited dynamic range: if part of the scene is extremely bright, the camera must adjust its exposure, causing the darker regions, representing the astronomical signals in this analogy to lose detail.

These limitations underline why radio telescopes are vulnerable to even modest interfering signals and why careful coordination with satellite operators is essential to preserve the dynamic range and integrity of scientific observations.

\subsection{Radio observatories and operations}

Major radio observatories are located in remote areas chosen for their natural radio quietness: high plateaus, deserts, and sparsely populated regions where terrestrial transmitters are few. Examples include the Karoo region in South Africa, the Murchison Radio-astronomy Observatory in Western Australia, and high-altitude valleys in America and Europe. In some cases, these observatories are surrounded by formal radio quiet zones (RQZs) in which terrestrial transmitters are restricted or subject to mandatory coordination. A prominent example is the United States National Radio Quiet Zone (NRQZ), established in 1958, which spans approximately 34\,000~km$^{2}$ across West Virginia, Virginia, and Maryland. Within this zone, fixed transmitters must be coordinated with the National Radio Astronomy Observatory (NRAO), and power levels or antenna orientations may be modified to limit interference to the Green Bank Telescope.

\begin{figure}[t]
    \centering
    \begin{subfigure}[b]{0.3\textwidth}
        \centering
        \includegraphics[width=\textwidth]{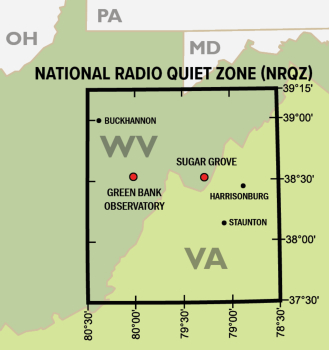}
        \caption{Map of the National Radio Quiet Zone (NRQZ).}
        \label{fig:rqz_map}
    \end{subfigure}
    \begin{subfigure}[b]{0.48\textwidth}
        \centering
        \includegraphics[width=\textwidth]{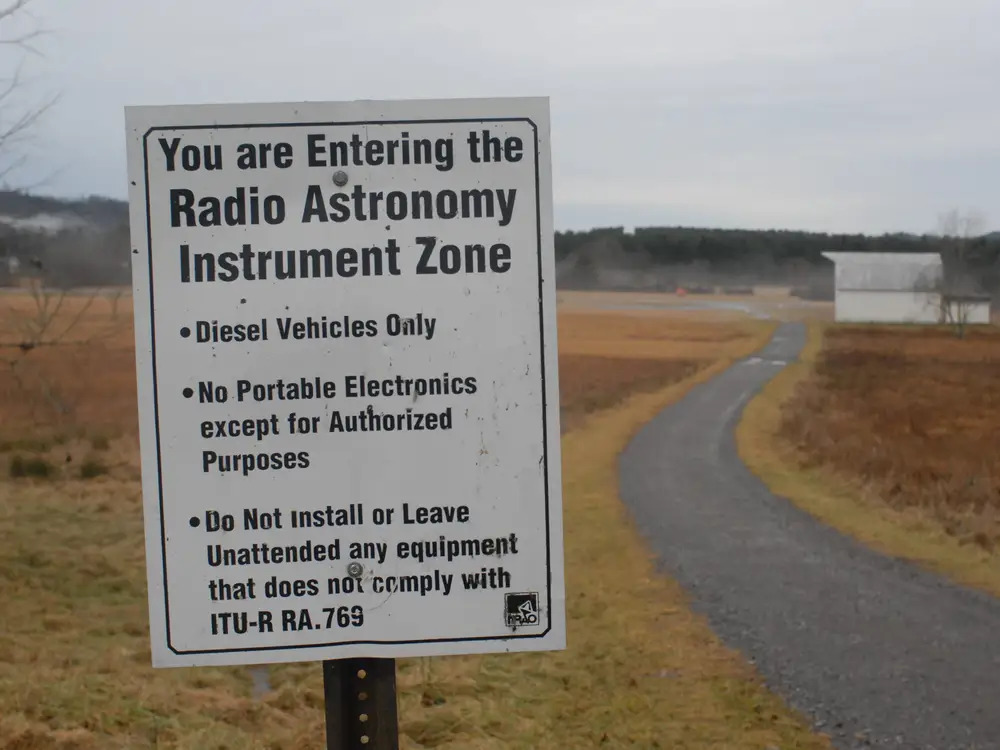}
        \caption{Sign indicating the entrance of the radio observatory within the NRQZ}
        \label{fig:rqz_sign}
    \end{subfigure}

    \caption{The United States National Radio Quiet Zone (NRQZ). Left: Map of the designated NRQZ, a federally managed region of approximately 34,000 km² spanning parts of West Virginia, Virginia, and Maryland. Within this zone, fixed terrestrial transmitters are subject to coordination to protect the Green Bank Observatory and the Sugar Grove facility (credit : NSF/NRAO). Right: A roadside sign marking the entrance to the Radio Astronomy Quiet Zone near Green Bank, West Virginia, indicating restrictions on the use of radio‐emitting devices to preserve the electromagnetic environment for scientific observations (credit : NSF/NRAO).}
    \label{fig:rqz}
\end{figure}

These zones are effective at suppressing ground-based emissions, but they cannot eliminate all sources of interference. Their jurisdiction applies only to terrestrial transmitters, not to aircraft, drones, or spacecraft. Consequently, satellite downlinks, out-of-band emissions, and unintended electromagnetic radiation from spacecraft subsystems remain major contributors to the interference environment even in the quietest locations on Earth.

Radio astronomy operations are diverse and scientifically driven. Some observations focus on extremely narrow spectral lines, requiring high spectral purity and resolution. Others rely on wide instantaneous bandwidths to study continuum emission or broad spectral features. Many modern facilities operate large-area survey programs in which the telescope continuously scans the sky, integrating briefly on each region. Time-domain studies, such as pulsar timing or searches for fast radio bursts (FRBs), require high time resolution and continuous monitoring of specific directions. In all of these modes, the requirements for spectral purity, time stability and dynamic range are demanding, and even weak, intermittent interference, can have outsized effects.

Calibration is central to all of these operations. Telescopes regularly observe well-characterized sources to measure their gain, phase, system temperature, and bandpass response. Interference complicates this process at several levels. Strong RFI can bias calibration measurements, introducing residual errors that propagate into science data, even when the interference is not visibly obvious. Some types of RFI can mimic genuine astronomical signals, producing false positives in transient searches, particularly when the interference exhibits frequency drifts or Doppler-like patterns that resemble the propagation signatures of astrophysical bursts. Conversely, strong or rapidly varying interference can mask short-duration astronomical events, effectively hiding real signals within corrupted data and reducing the scientific return of time-domain observations. Automated flagging removes contaminated time--frequency samples, but this comes at the cost of reduced effective integration time, which directly degrades the sensitivity of the observation, and can distort the recovery of extended or low-surface-brightness emission. Adaptive filtering and spatial nulling algorithms can help suppress certain types of interference, but they typically rely on assumptions about stationarity and may struggle with rapidly varying signals from low-Earth-orbit satellites.

Traditionally, radio astronomers have also taken advantage of ``spectrum holes'', which are relatively quiet frequency ranges lying between strong active services. Below 30~GHz, less than 2.5\% of the spectrum is allocated internationally to the Radio Astronomy Service (RAS), with only about 1.3\% allocated exclusively to passive use and 1.2\% allocated on a shared primary basis. Observing outside these protected bands is not optional: modern science goals, including wide-area surveys and cosmological experiments, cannot be accomplished within the limited set of protected passive allocations. As more services occupy more spectrum with increasingly complex and dense deployments, these natural gaps become narrower and less reliable. Dynamic spectrum sharing approaches, such as those discussed in the United States National Spectrum Strategy\citep{CRS-NationalSpectrumStrategy2020,WhiteHouse-SpectrumMemorandum2018}, raise additional challenges by increasing temporal occupancy, thereby reducing opportunities for passive services to exploit moments of quiet.

These trends contribute to a growing concern that without careful coordination, regulatory predictability, and technical mitigation strategies, certain science cases may become impractical or impossible. Ensuring long-term coexistence requires active engagement between observatories, administrations, and satellite operators, along with recognition that the passive use of the spectrum is both uniquely fragile and scientifically indispensable.

\section{Sensitivity and Interference}

\subsection{Radiometer equation and scaling}

The radiometer equation encapsulates how the sensitivity of a radio telescope improves with increasing bandwidth, integration time, and collecting area. In a simplified form, the uncertainty in the measured flux density \(S\) can be written as
\[
  \sigma_S \approx \frac{2 k_{\mathrm{B}} T_{\mathrm{sys}}}{A_{\mathrm{eff}} \sqrt{n_{\mathrm{pol}} \, \Delta \nu \, \tau}},
\]
where \(k_{\mathrm{B}}\) is Boltzmann's constant, \(T_{\mathrm{sys}}\) the system temperature of the receiver (the intrinsic noise floor of the instrument in the absence of astronomical signals or RFI), \(A_{\mathrm{eff}}\) the effective collecting area, \(n_{\mathrm{pol}}\) the number of polarizations, \(\Delta \nu\) the observing bandwidth, and \(\tau\) the total integration time. In an interferometric array with \(N\) antennas, the number of independent baselines is \(N(N-1)/2\), and the effective sensitivity improves roughly as the square root of this number, provided all baselines contribute independent measurements.

A key reason the radiometer equation works so well in radio astronomy is that most astronomical radio emission is intrinsically broadband. Synchrotron radiation, free--free emission and blackbody-like spectra all produce signals spread over wide frequency ranges, so integrating over large bandwidths is both natural and scientifically appropriate. Even narrow spectral features such as the H\,\textsc{i} 21-cm line or astrophysical masers are effectively broadened by cosmological redshift: distant galaxies shift these lines by tens or hundreds of megahertz, and wide-band receivers are required to search for them across large redshift ranges.

An important implication of the radiometer equation is that sensitivity scales as the square root of bandwidth. Any loss of usable bandwidth due to interference therefore has a direct and unavoidable impact on sensitivity. Because radio telescopes are engineered so that \(T_{\mathrm{sys}}\) is as low and as stable as possible, the only viable compensation for lost bandwidth is to increase the integration time \(\tau\). However, increasing \(\tau\) reduces the efficiency of the facility by leaving less observing time for other science programs. For heavily subscribed observatories, this loss of observing efficiency can propagate into significant scientific and operational costs.

\paragraph{Example: sensitivity loss from RFI flagging.}
The impact of losing bandwidth can be illustrated with a simple example. Suppose an observation uses a \SI{1}{\giga\hertz} bandwidth and achieves a root-mean-square noise level \(\sigma\). If \SI{10}{\percent} of the time--frequency pixels must be flagged due to interference, the effective bandwidth becomes
\[
  \Delta \nu_{\mathrm{eff}} = 0.9~\mathrm{GHz}.
\]
Because the radiometer equation predicts that sensitivity scales as \(1/\sqrt{\Delta \nu}\), the new noise level becomes
\[
  \sigma' = \frac{\sigma}{\sqrt{0.9}} \approx 1.054\,\sigma.
\]
Thus, losing only \SI{10}{\percent} of the bandwidth results in a degradation of sensitivity by about \SI{5.4}{\percent}. If \SI{30}{\percent} of the bandwidth is lost, a typical situation in heavily contaminated bands, the sensitivity penalty becomes
\[
  \sigma' = \frac{\sigma}{\sqrt{0.7}} \approx 1.195\,\sigma,
\]
corresponding to a \SI{19.5}{\percent} loss in sensitivity. Achieving the original sensitivity would require increasing the integration time by the same factor, i.e. by approximately \SI{20}{\percent} in this example. For highly subscribed facilities, this translates directly into reduced observing efficiency and fewer available hours for other scientific programs.

These considerations become especially acute for next-generation arrays with thousands of antennas observing over broad bandwidths for many hours. Such instruments can reach sensitivities where harmful interference corresponds to exceedingly small power spectral densities. In these regimes, weak out-of-band emissions (OOBE), unintended electromagnetic radiation from spacecraft subsystems, or low-level terrestrial signals, all of which may be invisible in short snapshots, can accumulate coherently or statistically across time and baselines, and significantly degrade the final data products.

Finally, it is worth noting that interference subtraction remains an active field of research. Many signal processing approaches have been proposed, including adaptive cancellation, matrix-based subtraction, sparse reconstruction and, machine-learning methods. However, none of these techniques has yet been deployed successfully in a real telescope for routine scientific observations. The primary reason is that even after sophisticated subtraction, residuals left behind by the processing chain remain orders of magnitude stronger than the astronomical signals of interest. As a result, robust scientific recovery of contaminated data has not yet been demonstrated, and prevention of interference remains vastly more effective than post-facto correction.

\subsection{Types of interference}

In radio astronomy, the term ``interference'' is used in a broad technical sense: it refers to any man-made emission that is detectable within the frequency range covered by a radio telescope. Importantly, this does not imply that the transmitter is violating regulations. Most interference that affects radio astronomy originates from entirely lawful services operating in their allocated bands. Legal interference becomes an issue only when emissions encroach upon bands allocated to the Radio Astronomy Service, but astronomers must regularly contend with detectable signals far outside those protected allocations.

The most straightforward case is in-band interference, which are emissions whose frequencies fall directly within the observing band of a telescope. Because telescopes make use of spectrum holes and shared bands to reach the sensitivity required by modern science, they are not restricted to bands allocated exclusively to passive services. If a satellite or terrestrial downlink transmits in or near a receiver's passband, its carrier and modulation spectrum can overlap with the broad natural astronomical emissions. Strong in-band signals can saturate low-noise amplifiers or analog-to-digital converters, driving them into non-linear regimes and rendering observations impossible. Even when filters are present in the signal chain, they come with compromises: each additional filter stage adds thermal noise, reduces dynamic range and increases system temperature, while also adding substantial cost for observatories deploying hundreds or thousands of receivers.

Out-of-band and spurious emissions are subtler but frequently more significant for radio astronomy. No transmitter is perfectly band-limited: modulation schemes create spectral side lobes, practical filters have finite roll-off, antennas exhibit frequency-dependent patterns, and power amplifiers introduce non-linearities. These effects generate out-of-band emissions (OOBE) that can extend well beyond the nominal service band. Regulatory emission masks typically set OOBE limits with reference to compatibility among communication systems, which are many orders of magnitude less sensitive than radio telescopes. As a result, emissions that are harmless for ordinary services may be detectable or even disruptive for passive scientific receivers.

Harmonics and intermodulation products represent another important class of unwanted signals. When strong signals pass through a non-linear device, they can interact to produce new frequencies at sums and differences of the originals. ``Strong signals interacting in shared hardware'' refers to situations where multiple transmitters on a spacecraft share amplifiers, power distribution networks, antennas, filters or mechanical structures. If any component in this chain exhibits non-linear behavior, even mildly, the signals can mix and generate tones at unexpected frequencies, some of which may fall within radio astronomy bands.

Passive intermodulation (PIM) is a related phenomenon that arises not from active electronics but from mechanical or structural elements that behave non-linearly under high current or intense electromagnetic fields. Seemingly innocuous components (bolted joints, oxidized connectors, dissimilar metals, loose fasteners or even antenna support structures) can form microscopic diode-like junctions. These act as weak mixers: when illuminated by strong transmissions, they generate low-level intermodulation products across a wide frequency range. Although the resulting PIM signals are far weaker than the primary transmissions, they can be well above the detection thresholds of radio telescopes.

Wideband noise emissions, whether intentional (e.g.\ spread-spectrum modulations) or unintentional (e.g.\ from switching power supplies, onboard computers or high-speed digital electronics), raise the noise floor over broad spectral regions. Bursty interference, typical of low-Earth-orbit satellites passing through the primary beam, or of GNSS satellites used in geolocation constellations, causes intermittent contamination that is particularly damaging to time-domain science and calibration processes.

Finally, aggregate interference from satellite constellations is a growing concern. A single spacecraft may be engineered so that its contribution at a given telescope is below the harmful interference thresholds defined in ITU-R Recommendation~RA.769 \citep{ITUR-RA769}. However, when hundreds or thousands of satellites are active simultaneously, each contributing a small amount of power, their combined power flux-density can exceed protective limits. Overlapping beams, complex duty cycles, and additional contributions from user terminals can exacerbate the problem. Accurately assessing these aggregate effects requires realistic modeling of orbits, antenna patterns, and operating modes across the entire constellation.

\subsection{Detection limits and RFI identification}

A fundamental reason radio telescopes can detect extremely faint interference lies in the way interferometers measure the sky. Rather than relying solely on the total power received by a single antenna, they compute cross-correlations, known as ``visibilities'', between pairs of antennas. Cross-correlation is sensitive only to signals that are coherent between antennas. Random thermal noise is uncorrelated from one receiver to another and therefore averages toward zero, while any coherent signal, such as a transmission from a satellite or aircraft, remains correlated and appears as a structured pattern in the visibilities or in the reconstructed images. This allows interferometers to extract coherent signals at levels far below the thermal noise floor of individual receivers. An interfering signal that is ``invisible'' in a single antenna’s power spectrum may nonetheless be detectable once correlated across an array.

Astronomers therefore use a variety of tools to detect, diagnose and mitigate radio frequency interference. Time--frequency representations (spectrograms) reveal narrowband carriers, drifting tones and impulsive bursts. Statistical tests compare the data against the Gaussian statistics expected from thermal noise, flagging outliers or non-Gaussian behavior. Spatial filtering techniques, such as forming interferometric images or direction-of-arrival estimates, can identify localized sources of interference and, in some cases, attempt partial subtraction. However, all such methods rely on assumptions about stationarity, stability and the separability of astronomical and anthropogenic signals. When interference is weak, intermittent or spectrally complex, the algorithms may either remove too much data, and therefore reducing sensitivity, or leave behind low-level residuals that still bias scientific results.

For satellite operators, the critical point is that radio astronomers routinely detect interference at levels far below the thresholds relevant to communication services. This is not a matter of unreasonable expectations but a direct consequence of the coherence-based techniques that enable radio astronomy to study extremely faint cosmic signals. Even emissions that appear negligible by conventional engineering standards may remain detectable, and sometimes disruptive, for modern radio telescopes.

\section{Why Satellites Matter to Radio Astronomy}

\subsection{LEO characteristics relevant to RAS}

Low Earth orbit (LEO) satellites, typically defined as spacecraft operating at altitudes between roughly \SI{160}{km} and \SI{2000}{km}, move rapidly across the sky as viewed from a ground-based radio telescope. Depending on altitude and the geometry of a pass, their apparent angular speed can range from one to several degrees per second near zenith, and somewhat slower at low elevations. This rapid motion determines how long a satellite remains within the main beam of a radio telescope and also governs the pattern of its traversal through the telescope's sidelobes. The associated Doppler shifts can be substantial: as the satellite approaches and recedes, the apparent frequency of narrowband emissions drifts noticeably, producing slanted or curved tracks in time--frequency spectrograms.

The response of a radio telescope is characterized by its beamwidth, typically defined by the full width at half maximum (FWHM) or the ``3\,dB'' width of the main lobe, i.e. the angular radius within which the received power drops to half of its peak value. For a single-dish antenna, the beamwidth is approximated by the diffraction-limited relation
\[
  \theta_{\mathrm{FWHM}} \approx 1.2\, \frac{\lambda}{D},
\]
often referred to as the \emph{Airy-disk} or \emph{diffraction-limited beamwidth} formula, where \(D\) is the diameter of the dish and \(\lambda = c/\nu\) is the observing wavelength expressed as a function of frequency \(\nu\) (with \(c\) the speed of light). A narrower beam implies a shorter main-beam crossing time for a fast-moving LEO satellite; for large dishes at high frequencies, this may be a fraction of a second. Nevertheless, satellites can remain detectable far outside the main lobe, since sidelobe gains, though much lower, still exceed the extraordinarily faint levels of astronomical signals. In interferometric arrays, different antennas or baselines may see a satellite in different sidelobes simultaneously, producing a complex pattern of correlated signatures.

The motion of satellites also complicates static mitigation strategies. Fixed notch filters, for example, remove interference only at a constant frequency, but a LEO satellite’s apparent transmission frequency can drift rapidly due to Doppler shifts. A fixed filter may therefore miss significant portions of the interfering signal as it sweeps across the band. Similarly, sky-exclusion zones or static avoidance regions are ineffective when spacecraft move quickly through the field of view, appearing and disappearing on timescales much shorter than typical astronomical integrations. Effective mitigation requires strategies that account for this rapid temporal, spatial and spectral variability, emphasizing the need for cooperation between satellite operators and radio observatories.

\paragraph{Example: main-beam crossing time.}
As a concrete example, consider a \SI{15}{m} dish observing at \SI{1.4}{\giga\hertz}, corresponding to a wavelength \(\lambda \approx \SI{0.21}{m}\). The diffraction-limited beamwidth is then
\[
  \theta_{\mathrm{FWHM}} \approx 1.2\,\frac{\lambda}{D} \approx 1.2\,\frac{0.21}{15} \,\mathrm{rad} \approx 0.017~\mathrm{rad} \approx 1^{\circ}.
\]
A typical LEO satellite can move across the sky at an apparent angular speed of order \(1^{\circ}\,\mathrm{s^{-1}}\) near zenith, so the time it spends within the main lobe of such a telescope is of order one second. Despite this short main-beam crossing time, the same satellite can remain detectable in sidelobes for tens of seconds before and after the main-beam transit, and in an interferometric array its signature can be seen across many baselines even when it is far from the pointing center.

\subsection{Aggregate constellation effects}

The primary reason satellites matter so acutely to radio astronomy today is not that any single spacecraft is exceptionally bright, but that their numbers are large and growing. Modern broadband constellations may include thousands of spacecraft, each transmitting continuously or quasi-continuously. Even if every satellite operates strictly within its allocated band and complies with individual power flux-density limits at the Earth's surface, the combined (or aggregate) contribution of many satellites can exceed the extremely low thresholds relevant for passive scientific receivers. For radio astronomy, where harmful interference limits correspond to received powers as low as \SI{-250}{dBW} or below, the cumulative effect of many individually compliant transmitters can easily become significant.

The concept of equivalent power flux-density (EPFD) is used in the ITU Radio Regulations to characterize this aggregate behavior. EPFD accounts for the instantaneous positions of satellites, their time-varying pointing directions, antenna gain patterns and duty cycles, and the angle-dependent gain of the victim radio astronomy antenna. The resulting EPFD value is therefore both time- and direction-dependent. EPFD limits are intended to ensure that non-geostationary systems do not impair the operation of passive services, including the Radio Astronomy Service (RAS), beyond specified protection thresholds. In practice, however, verifying EPFD compliance and relating formal EPFD studies to real-world observations can be challenging. Differences between assumed and actual antenna patterns, evolving deployment architectures and operational changes can all alter the interference environment.

Recent case studies illustrate that the problem is practical, not merely theoretical. A well-known example is the unwanted out-of-band emissions from the GLONASS navigation system near \SI{1612}{MHz}, a protected radio astronomy band used for observations of the hydroxyl (OH) spectral line. These emissions caused significant interference to radio observatories worldwide in the 1990s and early 2000s and required coordinated international mitigation efforts involving both operators and the astronomy community~\citep{glonass_mitigation_itu,glonass_rfi_nrao}. Similarly, interference associated with GNSS and other satellite downlinks has been reported near frequencies used for both radio astronomy and Earth exploration sensors, prompting investigations by national administrations and discussions within ITU-R Working Parties. These examples demonstrate that compatibility must be continuously evaluated as systems evolve, deployments scale, and aggregate effects grow in importance.

\subsection{Out-of-band emissions}

Most communication payloads are designed to operate within a defined allocated band. However, as discussed earlier, no practical transmitter is perfectly spectrally confined. Out-of-band emissions (OOBE) arise from several physical and engineering mechanisms: finite filter roll-off, modulation side-lobes, and various forms of non-linear distortion such as harmonics, intermodulation products and spectral splatter generated in high-power amplifiers or other non-linear components. Additional mixing and coupling effects within the spacecraft, for example in shared power amplifiers, multiplexers, or antenna feed networks, can also distribute energy into frequencies outside the intended transmission band. When a transponder operating near a radio astronomy band produces OOBE that intrudes into a passive allocation, it can degrade the performance of observatories, even when the transmitter remains compliant with its own service allocations.

The relevance of this issue for radio astronomy is underscored by WRC-27 Agenda Item~1.16, which calls for studies of unwanted emissions from non-geostationary satellite systems into primary radio astronomy bands and Radio Quiet Zones.\footnote{See Resolution~681 (WRC-23): ``Studies of technical and regulatory provisions necessary to protect radio astronomy ... from aggregate radio-frequency interference caused by systems in the non-geostationary-satellite orbit'' \citep{ITU-WRC23-Res681}.} These studies recognize that the proximity of active satellite downlink bands to sensitive radio astronomy allocations creates additional risk: even small amounts of unwanted emission can raise the noise floor of a passive observation.

Several design choices strongly influence the level of OOBE. The linearity of power amplifiers determines the degree of distortion and spurious emissions. The choice of modulation scheme affects the steepness of the spectral roll-off and the magnitude of side-lobes. The quality of analogue filters, the implementation of digital predistortion (DPD), and the degree of isolation between components all determine how effectively emissions are confined to the intended band. Power control algorithms, which increase transmit power under adverse link conditions, may unintentionally drive hardware closer to non-linear operation, increasing spurious emission levels. Mispointing, miscalibration or faulty terminals can generate unexpected spectral features or shift beams such that emissions illuminate sensitive sites more strongly than intended. User terminals and terrestrial gateways operating on the ground can also be significant sources of OOBE, especially when deployed within line-of-sight of observatories.

For radio astronomy, OOBE can be especially serious because protected passive bands often lie immediately adjacent to active service bands. For example, a band allocated to the Radio Astronomy Service for hydrogen line studies may sit directly beside a downlink band for a broadband satellite system. If filters are not sufficiently sharp, or if non-linearities generate substantial distortion products, the raised noise floor in the RAS band can undermine observations. Because radio astronomers integrate weak signals over long periods, even a modest and apparently insignificant increase in the noise floor can render faint astronomical signals undetectable.

\subsection{Unintended electromagnetic radiation}

Unintended electromagnetic radiation from satellites deserves special attention. This category includes all emissions that do not arise from the main communication transmitter, but from other subsystems such as onboard computers, clock distribution networks, reaction wheels, solar array regulators, star trackers, and payload electronics. Many of these systems operate with high-speed digital signals and switching power supplies. Unless great care is taken in electromagnetic compatibility (EMC) design, they can produce broadband and line emissions that couple to harnesses, structures and antennas and leak into space.

\begin{figure}[t]
    \centering
    \includegraphics[width=\textwidth]{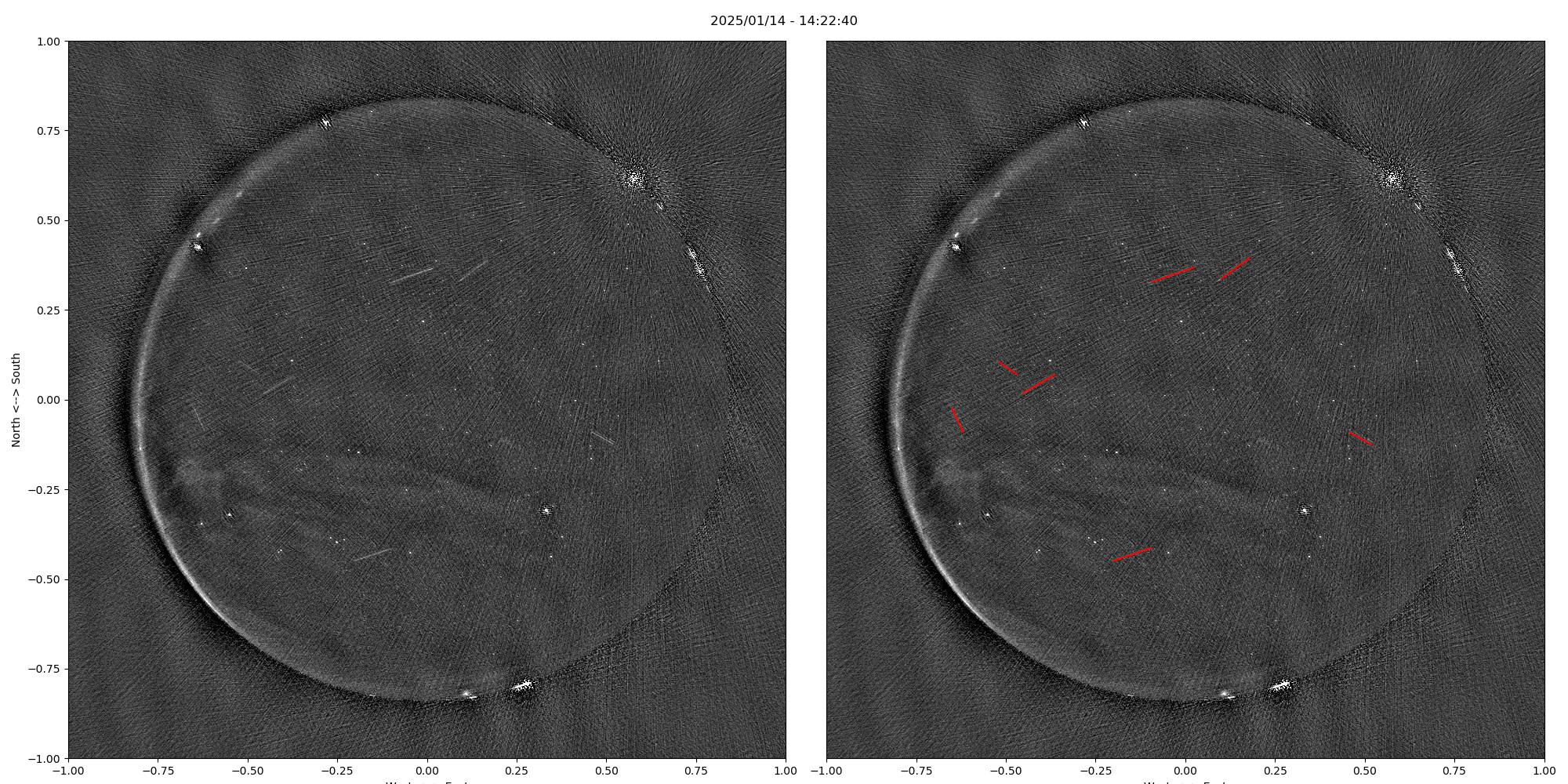}
    \caption{Example of a raw all-sky image produced with the OVRO–LWA telescope in California, USA (see photo \ref{fig:lwa}), in the 59–64 MHz band (not allocated to satellite services) during a 10-second observation on 2025-01-14. Left: Full-sky map in zenith-projection, where the circular boundary corresponds to the horizon and the center to the local zenith. Bright white spots along the edge indicate terrestrial transmitters, while several linear streaks across the sky correspond to satellite passes. Right: The same image, but with satellite streaks overlaid and identified using Two-Line Element (TLE) orbital data for known satellites in red.}
    \label{fig:example7}
\end{figure}

Radio telescopes are capable of detecting such emissions when satellites pass overhead, sometimes as broad spectral humps, and sometimes as narrow lines at clock harmonics. Because the emissions may not be well characterized, they can appear at unexpected frequencies, including those used by radio astronomy. Figure \ref{fig:example7} shows an example of such detection with the OVRO-LWA telescope. From the operator's perspective, they may be invisible, they do not affect the performance of the communication payload and may not be covered by standard tests focused on intentional emissions.

Managing unintended radiation is therefore an area where cooperation is both necessary and promising. Spacecraft builders can extend their EMC practices to consider the sensitivity of radio astronomy, using test campaigns that measure radiated emissions across wider frequency ranges and at lower levels than usual. Operators can share information with observatories about the frequencies and characteristics of any residual emissions that cannot be fully eliminated. Together, they can develop strategies to monitor and mitigate such effects in orbit.

\subsection{Atmospheric disturbances from launches and re-entries}

Beyond direct radio emissions, satellite activities can affect the propagation medium itself. Rocket launches and satellite re-entries inject energy and material into the upper atmosphere and ionosphere. Exhaust plumes and shock-acoustic waves can create transient depletion or enhancements in electron density, sometimes referred to as ``ionospheric holes'' or traveling ionospheric disturbances. Historical and recent studies have reported significant perturbations to ionospheric structure following major launches, including daytime depletion that persist for hours and wave-like disturbances detected in total electron content, and very-low-frequency (VLF) propagation measurements \citep{Bernhardt1976,Zinn1980,Feng2025Rocket}.

For most communication services, these effects manifest as modest changes in propagation delay, scintillation, or temporary degradation of certain links. For high-sensitivity, wide-field radio astronomy, the implications may be subtler but potentially important. At low and mid radio frequencies, especially below a few hundred megahertz, the ionosphere acts as a refracting and phase-changing layer in front of the telescope. Calibration strategies for cosmology and wide-field surveys often assume that the ionosphere varies smoothly in time and space and that its behavior can be modeled as a slowly changing phase screen. If rocket-induced disturbances introduce sharp gradients, ducts, or localized depletion, they may violate these assumptions.

Similarly, some cosmological experiments aim to detect extremely faint, spectrally smooth signals, such as the global 21-cm signature from the early Universe, or tiny fluctuations in the cosmic microwave background. These experiments rely on precise modeling of the instrument response and of the propagation effects of the atmosphere and ionosphere. Transient plasma structures associated with launches or re-entries could, in principle, imprint additional spectral or angular structure on the received sky signal, complicating foreground subtraction or calibration.

At present, there is no robust evidence that rocket launches or satellite re-entries have measurably biased major cosmological radio datasets. The number of high-cadence, high-sensitivity surveys operating simultaneously with intense launch campaigns has been limited, and many potential perturbations may be below current detection thresholds. However, the cadence of launches is increasing, and new studies are beginning to document the statistical properties of rocket-induced ionospheric disturbances with greater precision \citep{Feng2025Rocket,Chen2025Iono}. This has prompted interest in the radio astronomy community in systematically assessing whether such disturbances could affect future wide-field surveys, particularly at low frequencies.

From the perspective of satellite operators and launch providers, this emerging issue illustrates that the environmental footprint of space activities includes not only debris and emissions from radio transmitters, but also transient modifications of the propagation medium. As knowledge grows, there may be opportunities to coordinate particularly sensitive observations with launch schedules, or at least to ensure that data taken during known disturbance events are flagged and studied with appropriate models. For regulatory bodies and the United Nations system, these questions fit naturally into broader discussions of space sustainability and of the cumulative impact of human activities on near-Earth space.

\section{Regulation and Standards}

\subsection{The ITU system}

The International Telecommunication Union, a specialized agency of the United Nations, manages the global allocation and use of the radio-frequency spectrum. The ITU Radiocommunication Sector (ITU-R) maintains the Radio Regulations, an international treaty that defines frequency allocations to various radio services and sets conditions for their use. The Radio Astronomy Service (RAS) appears in many bands, sometimes on an exclusive basis, and sometimes on a secondary or shared basis. Footnotes associated with specific allocations often call attention to the need to protect radio astronomy observations.

Protection criteria for radio astronomy are set out in ITU-R Recommendation~RA.769, which defines levels of detrimental interference such that the long-term sensitivity loss in representative observing modes does not exceed ten per cent \citep{ITUR-RA769}. Other recommendations, such as RA.1513 and RA.1031, address protection in particular bands and coordination procedures \citep{ITUR-RA1513, ITUR-RA1031}. SM.329 and related texts discuss limits on unwanted emissions and sharing between services \citep{ITUR-SM329}. These documents provide a technical foundation for regulatory decisions taken at World Radiocommunication Conferences (WRCs).

A key concept embedded in these criteria is the harmful interference threshold. For a given telescope, observing mode, and band, RA.769 gives the interfering power at the receiver input that would cause a specified degradation in performance. To relate these thresholds to emissions from external sources, the interfering power can be converted into a corresponding limit on power flux-density (PFD) at the telescope site. PFD is a measure of how much electromagnetic power arrives per unit area per unit bandwidth at the location of the telescope, typically expressed in \(\mathrm{W\,m^{-2}\,Hz^{-1}}\) or in decibel units. It characterizes the strength of a signal as it reaches the ground, independent of the receiving antenna.

The conversion from a receiver-input power threshold to a PFD limit involves the telescope's gain pattern: the power collected by an antenna is the incoming power flux-density multiplied by the effective collecting area, which depends on the gain in the direction of the source. Because radio telescopes have extremely large effective areas and very high sensitivity, the resulting PFD thresholds are often many orders of magnitude below the levels relevant to communication systems. This reflects not regulatory strictness but the fundamental sensitivity of modern instruments operating near the limits imposed by physics.

\subsection{ITU-R Working Party 7D}

Within ITU-R, Study Group 7 deals with science services: radio astronomy, space research and Earth exploration. Working Party 7D is responsible for matters specifically related to radio astronomy. It prepares recommendations, reports and contributions to conferences on topics such as protection criteria, coordination zones, compatibility with emerging services, and the impact of new technologies.

Working Party 7D provides a forum where astronomers, national administrations, sector members, and industry representatives can discuss technical issues in a structured way. Satellite operators who engage with this process can help to shape realistic protection criteria and develop practical coexistence measures. Participation can also help operators anticipate future regulatory developments, including possible changes in allocations or in sharing conditions.

\subsection{National regulations}

National administrations implement the Radio Regulations and may impose additional protections for radio astronomy. In the United States, the National Telecommunications and Information Administration (NTIA) manages spectrum use by federal agencies, including observatories operated by federal institutions, while the Federal Communications Commission (FCC) regulates non-federal users. Licensing processes for satellite systems typically involve both bodies when federal incumbents may be affected. Coordination between operators and observatories can be formalized through conditions on licenses or through memoranda of understanding. The National Science Foundation (NSF), which supports and funds many of the nation's major radio observatories, plays a significant role in this process: NSF’s Spectrum Management Office represents the interests of its funded facilities in federal coordination proceedings, participates in NTIA committees, reviews proposed allocations or assignments that may affect scientific operations, and engages directly with the FCC when non-federal systems pose risks to NSF-supported observatories.

A further element of national implementation is the application of Article~4.4 of the ITU Radio Regulations. This provision states that administrations may authorize stations to operate in departures from the Radio Regulations, such as transmitting outside allocated bands, only under the condition that such stations do not cause harmful interference to services operating in accordance with the Regulations\footnote{ITU Radio Regulations, Article~4.4.}. In practice, Article~4.4 places the full burden of protection on the operator of the non-compliant or non-conforming system: they must not claim protection from interference, and they remain entirely responsible for avoiding harmful interference to compliant services, including the Radio Astronomy Service. National regulators often invoke Article~4.4 when authorizing experimental, temporary, or non-conforming operations, and radio observatories frequently rely on this provision when negotiating compatibility with nearby transmitters.

In other countries, spectrum regulators such as Ofcom in the United Kingdom or ANFR in France play analogous roles. Many administrations host national committees or working groups that bring together users from different sectors, including radio astronomy, to comment on proposed allocations and satellite filings. Some observatories benefit from national regulations that designate radio quiet zones, where terrestrial transmitters face stricter limits, building codes, or planning controls that help maintain a low-interference environment.

Satellite filings affect radio observatories in several ways. The core filing to the ITU includes technical parameters that are used in compatibility studies. National licenses may include conditions related to coordination with specified observatories. For operators, providing accurate technical data and engaging with concerns early in the process can prevent later objections and delays, and can significantly reduce the likelihood of interference disputes once the system is deployed.

\subsection{EPFD and its limitations}

The equivalent power flux-density (EPFD) is a quantity defined in the ITU Radio Regulations to assess the aggregate impact of non-geostationary satellite systems on protected services, including the geostationary fixed-satellite service and the Radio Astronomy Service. Conceptually, EPFD represents the total power flux-density arriving at a victim receiver from all satellites in view, weighted by the directional gain of that receiver in the direction of each satellite. It can be expressed, in a notional form, as
\[
  \mathrm{EPFD}(\theta, \phi) = 
  \sum_i \frac{P_i \, G_i(\theta_i,\phi_i)}{4\pi R_i^2} \, 
  G_{\mathrm{victim}}(\theta_i,\phi_i),
\]
where the sum is over satellites \(i\), \(P_i\) is their transmit power, \(G_i(\theta_i,\phi_i)\) is the satellite antenna gain toward the victim receiver (e.g.\ a radio telescope), \(R_i\) is the distance to the receiver, and \(G_{\mathrm{victim}}(\theta_i,\phi_i)\) is the gain of the victim antenna in the direction of that satellite. In practice, formal EPFD evaluations are more complex and typically involve statistical simulations over satellite orbital positions, time-varying beam pointings, modulation patterns and duty cycles.

EPFD compliance is usually demonstrated through Monte Carlo simulations following prescribed ITU methodologies, complemented in some cases by in-orbit measurement campaigns. However, for radio astronomy, several factors limit the extent to which formal EPFD compliance alone can guarantee protection from harmful interference. The EPFD limits in the Radio Regulations were derived under specific historical assumptions about antenna sizes, observing modes and operating environments that do not always reflect modern instruments, particularly widefield interferometers and arrays with thousands of elements. The antenna patterns used in compatibility studies may underestimate real sidelobe levels, especially when telescopes operate in scanning modes, or at frequencies where the beamshape deviates from idealized models.

Furthermore, EPFD calculations generally account only for in-band, intentionally radiated emissions. Unintended components such as out-of-band emissions, harmonics, intermodulation products, or emissions from user terminals and terrestrial gateways may not be fully represented in the models. These additional contributors can raise the effective noise floor or produce transient contamination even if the nominal EPFD limit is respected. As a result, while EPFD provides a useful regulatory framework for assessing aggregate interference, it is not a substitute for detailed, case-by-case technical analysis, and for sustained dialogue between satellite operators and radio observatories.

\subsection{The emerging EU Space Act}

Within the broader regulatory landscape, the European Union has recently proposed a new horizontal framework for space activities: the EU Space Act. Presented by the European Commission in June 2025, this proposed regulation aims to create a single market for space activities in the Union by harmonizing currently fragmented national regimes and establishing common rules on safety, resilience, and environmental sustainability for space services and infrastructure \citep{EC-EUSpaceAct2025,EUParlSpaceLaw2025}. It is intended to apply not only to operators based in EU Member States, but also to non-EU operators that provide space services within the EU.

The proposal rests on several pillars. The safety pillar introduces requirements related to space traffic management, collision avoidance, and debris mitigation. Operators would need to demonstrate that their missions meet common standards for end-of-life disposal, maneuverability, and risk management. The resilience pillar addresses cybersecurity and continuity of space-based services, recognizing that satellites are critical infrastructure for European economies and societies. The sustainability pillar requires operators to assess and reduce the environmental impacts of their activities, including effects on the space environment and, potentially, on the Earth's atmosphere and electromagnetic environment \citep{EC-EUSpaceAct2025,EUParlSpaceLaw2025}.

While the EU Space Act is not a spectrum-management instrument in the narrow ITU sense, it intersects with radio astronomy in several ways. First, its emphasis on safety and sustainability may lead to stronger expectations that operators identify and mitigate interference risks to scientific uses of space and spectrum as part of their environmental and risk assessments. Second, by creating a common authorization framework at EU level, it may streamline how conditions related to coexistence with radio astronomy are attached to licenses, reducing fragmentation among Member States. Third, the Act's provisions on monitoring and incident reporting could be used to structure how electromagnetic interference events, including those affecting observatories, are documented and addressed.

From the perspective of satellite operators, the EU Space Act signals that the European regulatory conversation is moving beyond narrow licensing towards a more integrated view of space as an environment that must be managed for safety, resilience, and sustainability. For radio astronomy, it offers an opportunity to embed protection of ``quiet skies'' and of the electromagnetic environment within a broader policy framework, alongside debris mitigation and climate-related considerations. The proposal is still under negotiation in the European Parliament and the Council at the time of writing, and its final form will depend on the outcome of those legislative deliberations. Nevertheless, its direction of travel is clear: space activities in Europe will increasingly be expected to demonstrate compatibility with long-term environmental, safety and scientific objectives, including the protection of the electromagnetic environment and ``quiet skies'' for passive services.

\section{Practical Coexistence for Satellite Operators}

For satellite operators, the most important question is what can be done in practice to coexist with radio astronomy. The answer spans both engineering measures and organizational processes, and successful coexistence typically requires a combination of both.

On the engineering side, managing power spectral density is central. Transmission schemes can be designed with conservative margins on out-of-band emissions, using high-quality analogue filters, carefully linearized power amplifiers, and modulation formats that minimize spectral expansion due to harmonics and intermodulation products. Digital predistortion can further improve linearity and suppress unwanted emission products. Antenna patterns can be engineered with particular attention to sidelobe levels in the directions of known radio observatory locations. Beam-pointing strategies may avoid directing high-gain beams at low elevation angles toward the horizon where they intersect sensitive sites, or reduce transmit power when satellites pass above major observatories at frequencies adjacent to protected passive bands.

\begin{figure}[t]
    \centering
    \includegraphics[width=\textwidth]{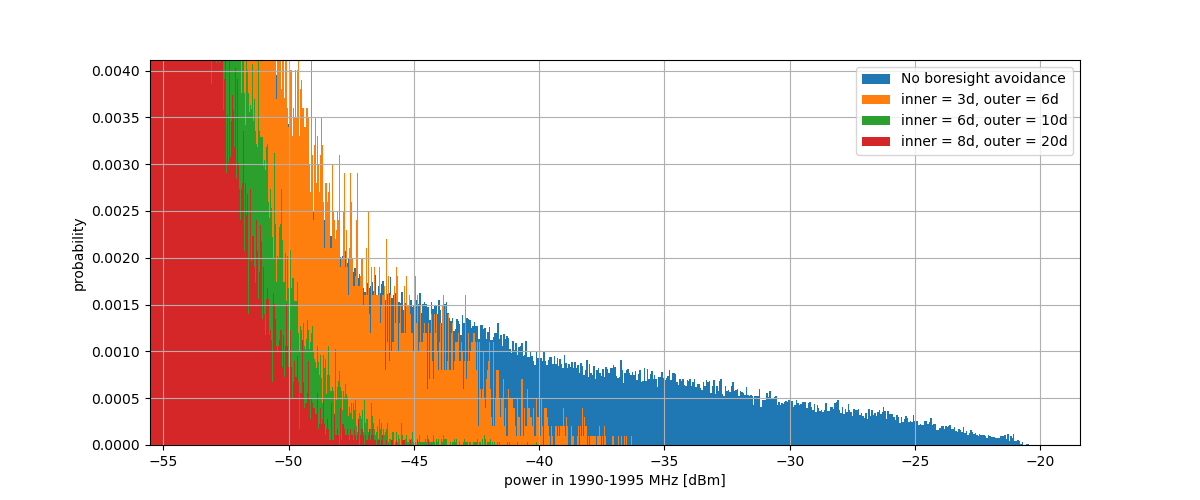}
    \caption{Impact of the increasing angle of boresight avoidance. Distribution of instantaneous powers collected with a DSA-2000 prototype antenna with multiple configuration of boresight avoidance (orange, green, red) compared to the powers collected without boresight avoidance. Up to 25 dB attenuation can be measured with the most conservative boresight avoidance configuration.}
    \label{fig:boresight_avoidance}
\end{figure}

Duty cycles and burst scheduling offer additional levers. Operators may choose to avoid test transmissions or particularly intense data bursts during periods when major observatories are conducting time-critical observations in vulnerable bands. Where feasible, dynamic avoidance schemes can be implemented that rely on knowledge of telescope schedules and instantaneous beam pointings. One increasingly discussed method is boresight avoidance, in which the precise instantaneous pointing direction of a radio telescope's main beam is shared with operators through a simple, secure Application Programming Interface (API). In return, satellites adjust their beams or power levels whenever their high-gain main lobe would intersect the telescope's boresight within a defined angular radius. This technique requires minimal data exchange but can substantially reduce the probability of harmful main-beam illumination events, as can be seen in figure \ref{fig:boresight_avoidance}.

Verification of out-of-band and spurious emissions remains a key element. Operators can perform laboratory and over-the-air tests to measure emissions across a wide frequency range, including passive bands used by radio astronomy. Sharing high-level results with observatories and regulators (for instance, spectral masks, measured attenuation levels at adjacent passive frequencies, and verification of compliance margins) builds transparency and allows astronomers to plan observations with realistic expectations and updated risk assessments.

On orbit, emissions may evolve over time. Component aging, thermal drift, and changing operational modes can alter emission characteristics, especially as hardware is pushed toward the edges of its performance envelope. Regular monitoring and, when necessary, in-orbit adjustments or firmware updates can help maintain compatibility. Faulty user terminals, misaligned Earth-station antennas, or malfunctioning spacecraft can become unexpected emitters. Having procedures in place to detect, diagnose, and remediate such anomalies promptly is in the interest of operators, regulators and the scientific community alike.

Coordination best practices revolve around communication and consultation. Proactive engagement with observatories before deployment allows potential compatibility issues to be identified early, while design decisions are still flexible. Data sharing, within commercial and security constraints, supports dynamic avoidance strategies and helps refine aggregate-interference models. Standardized procedures for reporting interference events, including agreed formats, time stamps, metadata and points of contact, greatly accelerate diagnosis. Joint test campaigns, in which operators and observatories coordinate transmissions, telemetry, and measurements, have historically resolved difficult interference problems. A notable example is the successful mitigation of unwanted emissions from GLONASS satellites near the \SI{1612}{\mega\hertz} hydroxyl line, achieved through coordinated analysis and technical collaboration between system operators and the astronomy community.

These measures are not only protective. They also help operators demonstrate due diligence and responsible behavior in the eyes of regulators, funding agencies, the public, and international bodies focused on sustainable space operations. By integrating coexistence considerations proactively, operators strengthen their credibility and reduce long-term regulatory and reputational risk.

\section{Future Directions}

Radio astronomy and satellite operations are both entering periods of rapid expansion. Next-generation facilities such as the DSA-2000, the Square Kilometre Array (SKA), the ngVLA, and other wide-field survey instruments will deliver order-of-magnitude improvements in sensitivity, time resolution, and survey speed. These facilities are designed to operate for several decades and will support long-term cosmological experiments, pulsar timing arrays, transient monitoring, and wide-field cartography of the radio sky.

In parallel, the number and diversity of satellites operating in near-Earth space is increasing rapidly. Constellations in low Earth orbit (LEO, below roughly \(2{,}000~\mathrm{km}\)) and medium Earth orbit (MEO, between \(\sim 2{,}000\) and \(35{,}786~\mathrm{km}\)) are expected to grow substantially, driven by global demand for broadband connectivity, Earth observation, sensing, and emerging commercial services. This densification of the orbital environment means that future radio telescopes will operate alongside far more (and far more varied) spaceborne transmitters than in previous decades.

\subsection{Very Low Earth Orbit (VLEO) and emerging deployment trends}

A particularly dynamic area of development is very low Earth orbit (VLEO), typically defined as altitudes below \(\sim 450~\mathrm{km}\) and, in some demonstration missions, as low as \(\sim 250\)–\(\sim 300~\mathrm{km}\). VLEO offers reduced latency for communications, higher spatial resolution for Earth observation and potential reductions in long-term debris risk due to rapid orbital decay. Several commercial and governmental actors are now testing VLEO-capable platforms, and future large constellations may include significant numbers of satellites in these regimes.

From the standpoint of radio astronomy, VLEO presents both benefits and challenges. Rapid orbital decay limits debris accumulation, yet the closer range of VLEO satellites increases the power flux-density received at Earth by a factor proportional to \(1/R^{2}\). Their angular speeds across the sky are higher than those of traditional LEO satellites, producing shorter but potentially more intense illumination of telescope beams. Shorter lifetimes also imply more frequent launches and re-entries, with secondary consequences for atmospheric conditions and calibration, as discussed in the next subsection.

\subsection{Atmospheric drag, re-entry plasma effects and calibration impacts}

High-sensitivity radio astronomy increasingly depends on accurate models of the signal propagation environment. At low and mid frequencies, the ionosphere and upper atmosphere impose refractive shifts, dispersive delays, and spatially variable distortions that must be calibrated with high precision to recover detailed sky maps and accurate fluxes. Arrays such as LOFAR, HERA, MWA, and future instruments like SKA-Low and DSA-2000 require exceptionally stable atmospheric conditions or sophisticated calibration pipelines.

Rocket launches, upper-stage engine burns, and the de-orbiting of satellites generate transient ionospheric perturbations, including plasma trails, localized density enhancements, shock-induced structures and chemical alterations that temporarily modify electron densities. While such disturbances have not yet been shown to routinely compromise radio astronomy data, they remain under active study, especially in the context of increasing launch cadence and the higher frequency of re-entry events associated with VLEO operations. As cosmological surveys, 21-cm experiments, global-signal measurements, and pulsar timing arrays push toward tighter calibration tolerances, even subtle atmospheric disturbances may become scientifically relevant.

\subsection{Dark and Quiet Skies and evolving policy expectations}

The global conversation on preserving the optical and radio environment has been strongly influenced by the ``Dark and Quiet Skies'' initiative, coordinated by the United Nations Office for Outer Space Affairs (UNOOSA), the International Astronomical Union (IAU) and the International Telecommunication Union (ITU). Reports issued by these expert groups \citep{DQSI, DQSII} synthesize scientific concerns and propose voluntary guidelines for mitigating the impact of satellites on optical brightness and radio-frequency interference.

Recommendations include controlling satellite albedo, adopting orbital configurations that reduce sky brightness, minimizing out-of-band emissions near passive service bands and establishing enhanced coordination mechanisms between satellite operators and observatories. Although non-binding, these guidelines carry growing normative weight. They now inform discussions at the UN Committee on the Peaceful Uses of Outer Space (COPUOS), national licensing decisions, and regional regulatory initiatives such as the emerging EU Space Act.

As environmental stewardship becomes a more explicit part of space governance, operators who anticipate these expectations by adopting lower-impact designs, committing to transparency, and developing coexistence strategies, will be better positioned in an international environment where responsible space behavior is increasingly valued.

\subsection{A shared scientific and orbital ecosystem}

Emerging technologies also offer opportunities for cooperation. Machine-learning–based interference classifiers, adaptive beamforming techniques, and joint scheduling systems that share telescope boresight information via secure APIs could play meaningful roles in future coexistence frameworks. On the regulatory front, new instruments may evolve that explicitly recognize radio astronomy as a global public good while maintaining flexibility for commercial operators.

Ultimately, sharing the radio sky responsibly is a collective enterprise. Radio astronomers must continue to articulate their requirements and progress mitigation research, even as satellite operators innovate and refine their emission control strategies. Regulators and international bodies, including those within the United Nations system, will play a critical role in ensuring that governance frameworks evolve in ways that balance innovation, commercial development and the preservation of humanity’s ability to study the Universe. The decisions made over the coming decade will shape not only the scientific landscape but also the sustainability of the shared electromagnetic environment on which both science and industry depend.

\section{About the author}

Dr Gregory Hellbourg is a radio astronomer and spectrum manager specializing in the protection of scientific observations from radio frequency interference. He serves as a Staff Scientist in the Physics, Mathematics and Astronomy Division at Caltech and at the Owens Valley Radio Observatory, where he leads interference monitoring and spectrum coordination activities for several major radio astronomy facilities.

His scientific background spans radio instrumentation, interference detection and mitigation, and the search for technosignatures, with prior work at UC Berkeley, and CSIRO and ICRAR in Australia. He has developed advanced techniques for identifying, characterizing and mitigating interference in both single-dish and interferometric observations, and has contributed to the understanding of how satellite constellations, aircraft transmissions, spectrum congestion and atmospheric variability affect sensitive radio measurements. His technical work includes the development of real-time RFI monitoring systems, autonomous hardware for spectrum surveillance, and new statistical flagging algorithms for interferometric arrays.

Dr Hellbourg plays an active role in the emergence of new radio facilities, including the Deep Synoptic Array (DSA-2000) project in Nevada, where he helps shape spectrum policy, quiet-zone planning, and the integration of interference-aware operations into the scientific pipeline. He also leads the development of an initiative to standardize how observatories document and analyze interference events across the global community.

In addition to his research activities, Dr Hellbourg has organized multiple international conferences, workshops and training schools devoted to advancing spectrum policy, RFI mitigation, and the protection of passive scientific services. He has led educational programs designed to broaden participation in spectrum management, support early-career scientists and engineers entering the field, and raise global awareness of the scientific importance of maintaining a quiet radio environment. These efforts contribute to the international capacity-building goals shared by scientific institutions and UN-affiliated organizations.

Beyond research, Dr Hellbourg works at the intersection of science, policy, and space governance. He contributes to national and international discussions on space sustainability, out-of-band emissions, coordination between satellite operators and observatories, and the protection of the electromagnetic environment as a scientific commons. He serves as a member of the United States delegation to ITU-R Working Party 7D, where he participates in global technical and regulatory processes related to radio astronomy and space science services. He also contributes to the United Nations Committee on the Peaceful Uses of Outer Space (COPUOS) as an expert member, engaging with international efforts to balance the rapid growth of space-based services with the long-term preservation of the radio sky.

Through his technical expertise, field experience and policy engagement, Dr Hellbourg advocates for cooperative, evidence-based solutions that allow satellite operators, regulators, and scientists to share the radio spectrum responsibly while preserving humanity’s ability to explore the Universe.

This work is supported by NSF grants \#2229497, \#2229428, \#2128497.

\newpage

\section*{Fact Sheet: Protecting Radio Astronomy in the Age of Mega-Constellations}
\addcontentsline{toc}{section}{Fact Sheet}

\subsection*{What makes radio astronomy uniquely vulnerable?}
Radio telescopes detect natural signals with flux densities as low as
\(10^{-29}~\mathrm{W\,m^{-2}\,Hz^{-1}}\), many trillions of times weaker than communication signals.  
A harmful interference threshold in a protected band may correspond to a received power of order \(\sim\SI{-250}{dBW}\) at the receiver input.  
Even extremely faint out-of-band or spurious emissions, invisible to conventional radiocommunication systems, can be detectable in modern interferometers after long integrations.

\subsection*{Key ITU concepts and protections}
\begin{itemize}\setlength{\itemsep}{1pt}\setlength{\parskip}{0pt}
    \item {Radio Regulations (RR)}: treaty-level allocations and conditions for all services.
    \item {Article 4.4}: non-conforming stations must not cause harmful interference and cannot claim protection.
    \item {Recommendation ITU-R RA.769}: defines detrimental interference thresholds for radio astronomy.
    \item {EPFD}: ``equivalent power flux-density'' used to evaluate aggregate NGSO interference.
    \item {RA.1031, RA.1513}: additional guidance on sharing and protection of passive bands.
\end{itemize}

\subsection*{Satellite characteristics of concern}
\begin{itemize}\setlength{\itemsep}{1pt}\setlength{\parskip}{0pt}
    \item {In-band transmissions}: can saturate LNAs, mixers or ADCs.
    \item {Out-of-band emissions}: caused by finite filter roll-off, spectral side-lobes, intermodulation, harmonics, or non-linear amplifiers.
    \item {Unintended emissions}: from power converters, digital electronics, clocks, harness coupling, or other payload subsystems.
    \item {Aggregate effects}: thousands of individually compliant satellites can exceed RAS limits collectively.
    \item {Rapid LEO motion}: produces Doppler sweeps across bands, complicating fixed notches and static mitigation.
\end{itemize}

\subsection*{Why VLEO matters?}
Very Low Earth Orbit (VLEO, typically below \(\sim450~\mathrm{km}\)) increases:
\begin{itemize}\setlength{\itemsep}{1pt}\setlength{\parskip}{0pt}
    \item Received power flux-density (\(\propto 1/R^{2}\)).
    \item Apparent angular speeds, reducing time for avoidance actions.
    \item Launch frequency and re-entry rate, with implications for ionospheric structure.
\end{itemize}

\subsection*{Atmospheric and ionospheric impacts}
\begin{itemize}\setlength{\itemsep}{1pt}\setlength{\parskip}{0pt}
    \item Rocket plumes, engine burns, and re-entries create transient plasma structures.
    \item Localized disturbances may affect calibration for low-frequency cosmology, global 21-cm experiments, and precision time-domain surveys.
\end{itemize}

\subsection*{Best practices for satellite operators}
\begin{itemize}\setlength{\itemsep}{1pt}\setlength{\parskip}{0pt}
    \item Maintain conservative out-of-band emission margins, and verify with wide-range EMC/RF tests.
    \item Use digital predistortion, linearized PAs, and high-quality analogue filtering.
    \item Control sidelobe levels and avoid illuminating observatories with high-gain beams.
    \item Implement {boresight avoidance}: use a secure API to receive telescope pointing information and temporarily adjust power or beam shape.
    \item Coordinate burst scheduling and high-power tests away from time-critical astronomical observations.
    \item Establish clear procedures for anomaly reporting, ephemeris sharing, and rapid interference investigations.
\end{itemize}

\subsection*{Why operators should care}
\begin{itemize}\setlength{\itemsep}{1pt}\setlength{\parskip}{0pt}
    \item {Regulatory stability}: compliance with RR, EPFD limits, and national requirements reduces licensing risk.
    \item {Reputation}: avoiding interference to major observatories demonstrates responsible behavior aligned with space sustainability expectations.
    \item {Operational resilience}: understanding interference pathways improves system robustness and EMC performance overall.
\end{itemize}

\subsection*{Global policy trends}
\begin{itemize}\setlength{\itemsep}{1pt}\setlength{\parskip}{0pt}
    \item UN ``{Dark and Quiet Skies}’’ studies highlight both optical and RF impacts.
    \item Environmental protection of the {electromagnetic environment} is emerging as a component of space sustainability.
    \item Regional initiatives (e.g.\ the {EU Space Act}) increasingly integrate interference management within environmental, safety and resilience frameworks.
\end{itemize}

\subsection*{Core message}
Radio astronomy and satellite operators share the same sky. Protecting scientific access to the radio spectrum requires a combination of technical diligence, transparent coordination, and modern regulatory tools. When cooperation is proactive, coexistence is not only possible but mutually beneficial, enabling both advanced scientific discovery and sustainable satellite services.

\end{document}